\documentclass[copyright,creativecommons]{eptcs}
\providecommand{\event}{FAVO 2009} 
\providecommand{\volume}{??}
\providecommand{\anno}{20??}
\providecommand{\firstpage}{1}
\providecommand{\eid}{??}
\usepackage{breakurl}        
\usepackage{graphicx}

\title{A Framework to Manage the Complex Organization of Collaborating: Its Application to Autonomous Systems}
\author{Peter Johnson \qquad \qquad Rachid Hourizi \qquad \qquad  Neil Carrigan \qquad \qquad  Nick Forbes
\institute{Human Computer Systems Group,  Department of Computer Science\\
School of Computer Science and Engineering,  University of Bath, Bath, UK}
\email{p.johnson@bath.ac.uk}
}
\def\titlerunning{The Organisation of Collaboration}
\def\authorrunning{Peter Johnson,  Rachid Hourizi,  Neil Carrigan, Nick Forbes}
\begin{document}
\maketitle

\begin{abstract}
In this paper we present an analysis of the complexities of large group collaboration and its application to develop detailed requirements for collaboration schema for Autonomous Systems (AS). These requirements flow from our development of a framework for collaboration that provides a basis for designing, supporting and managing complex collaborative systems that can be applied and tested in various real world settings. We present the concepts of ``collaborative flow'' and ``working as one'' as descriptive expressions of what good collaborative teamwork can be in such scenarios. The paper considers the application of the framework within different scenarios and discuses the utility of the framework in modelling and supporting collaboration in complex organisational structures.  
  \end{abstract}

\section{Introduction}

This paper addresses important conceptual issues concerning collaboration in groups and sets out a description of the nature of the complexities of large group collaboration. It provides a basis for thinking about the structural aspects of collaboration in Virtual Organisations from both a technical and social perspective. It describes succinctly how the findings may be applied to autonomous systems, based upon our past and current research on collaboration \cite{1,2} and autonomous systems \cite{3}. 
 Why should we be interested in both group and collaborative working for complex human and autonomous systems (AS)? Group working is rather obvious in that many activities will require more resources, capability and effort that a single AS cannot provide. The second is less obvious, why should we be interested in collaborative working? The nature of working in teams and managing groups is often harder than we realize, the costs of working together can at times outstrip the benefits. The question is what does it mean to work together? Simply bringing a collection of people or agents or machines together does not achieve ´working together¡. The difference between a group that works together and one that does not shows its effects in many ways including the quality and efficiency of the work, the ease of working in the group, the ease of managing the group and the level of confidence one can have in that unit. A group which works together well achieves a high-level of ´flow¡ to its work enabling a state of operation in which the individuals and the group as a whole are fully immersed in what they are doing by a feeling of energized focus, full involvement, and success in the process of the activity \cite{3}. 
To have a group (of people or AS) work together requires more than just enabling them to communicate and coordinate well, it requires collaboration. Collaboration requires both good communication and good coordination; it also requires that they work as one with shared goals, shared understanding, with a ´common ground¡\cite{4,5}. It does not necessarily require a ´leader¡ and can occur in different group structures \cite{6}. Collaborative groups can achieve flow and produce greater quality of product more efficiently \cite{7}, with easier group working and with easier management. Moreover collaborative groups can achieve ‘flow’ and produce greater levels of creativity \cite{8,9}. The aim of this paper is to understand how large groups can function as collaborative groups and how this needs to be adapted to be applied to particular settings, in this first instance AS.
Achieving this aim has required us to develop a conceptual understanding of the nature of large group collaboration, of the ways in which it is achieved and the benefits that accrue when that achievement takes place. In applying this to AS requires us to identify the particular characteristics of large group collaboration involving autonomous systems, the problems and challenges that can arise whilst it is taking place and the ways in which those problems and challenges can be overcome.
With respect to its application to deploying multiple AS the potential benefits are well documented \cite{3, 10}. These benefits include the availability of a greater range of resources than are possessed by any single AS (e.g. the ability to search an area that includes both aerial and underwater threats), the enhancement of mission completion (e.g. the speed of completion, parsimony of resource utilization and reliability of outcome) and the achievement of mission objectives that lie beyond the scope of any single system (e.g. the simultaneous screening of many thousands of people inside a crowded street or public space).
To take full advantage of these benefits we must develop systems that
can do more than co-exist. More specifically, we must design systems
that can coordinate their roles, objectives, data, resources and
activities in such a way as to achieve smooth, low cost work with
minimal disruptions and conflicts. We describe this smooth, efficient
multi-actor activity as "collaboration" and draw upon previous work
\cite{11} to understand its optimal operation as one involving multiple
actors ``working as one'' and achieving collaborative "flow"
\cite{12}. This smooth efficient collaboration is difficult to achieve,
even when group sizes are small (i.e. in groups of five or less,
co-located actors, pursuing clearly defined, well-understood tasks)
and the goals, actions, understandings and impact of other actors are
easy to identify. Achieving collaborative flow in situations that
require the involvement of larger sized and/or multiple-groups of
actors is yet more difficult. Each of those actors may have different
capabilities, pursue multiple goals and be involved in many different
activities. In this context, the designers, managers and participants
of large-group collaborations cannot rely upon the existence of shared
or common understanding, such as that which exists within smaller
groups \cite{6,13}.  
In smaller groups, each actor is often able to follow the goals, activity, tasks, resources and capabilities of each other actor (1:1 understanding) \cite{5}. In large groups, by contrast this 1:1 understanding is less prevalent (i.e. an actor may understand the goals, activities, tasks, resources and capabilities of some but not all other actors). In this context actors within large group collaborations may require strategies/mechanisms that allow them to develop and use more abstract or group-level understanding of each others goals, actions tasks etc \cite{6}, in addition to the more detailed understanding of these attributes that they may have of some subset of actors in the group. (Note: even in a large group there may still be some 1:1 understanding but it will not be developed between every actor in the group and every other actor).
Larger-scale collaborations are, as a result, both qualitatively and
quantitatively different from the small-group collaboration, in the
sense that the possibility for a variations and individual differences
in the goals, actions and understandings that exist within the group
may not be understood easily or in great detail by the group or its
members. Consequently, large-scale collaborations require both the
group(s) and its (their) members to manage the understandings of and
contributions to the multiple goals and activities. Moreover, in
highly dynamic situations in which the goals, resources, group members
etc are likely to be changeable or emergent there is even greater
complexity to the collaboration structures and processes. 

\subsection{Collaboration: the application to AS}

Our research (as part of the SEAS DTC- Systems Engineering for
Autonomous Systems Defence Technology Centre) has identified the
capabilities required of AS that enable participation in these
large-scale collaborations, considers the costs of deploying AS
without such capabilities, tests the benefits of deployments involving
collaboratively capable AS and demonstrates the effects of such
deployment in authentic scenarios of use. The research addresses four
questions crucial for effective AS collaboration: 

\begin{enumerate}
\item What mechanisms or strategies for coping do groups of AS need to achieve/maintain collaboration?
\item When AS have to collaborate with other AS within these large groups, what are the coping strategies/mechanisms that they require for this?
\item What coping mechanisms/strategies do AS have to initiate a request for collaboration within large groups? 
 \item What efforts are required by People to work with AS that either
   have or do not have those capabilities (i.e. what are the savings
   to the task and collaboration costs imposed upon those humans)?
\end{enumerate}

\subsection{Benefits of Collaborative AS}

The understanding gained by addressing these four questions has
enabled the identification of the functionality required of
next-generation autonomous systems, capable of managing their own
contribution to wider system goals and has the potential to deliver
the one human to multiple platform vision. This will provide for:  
\begin{itemize}
\item  increased ``Flow'' in the work undertaken by large groups of AS,
\item  improved coordination/reduced coordination problems within those groups,
fewer interruptions,
\item  fewer/less severe communication problems,
\item easier and more efficient management of large groups of AS,
\item improved quality of performance and product.
\end{itemize}
The research contributes to an understanding of the additional ``coping
strategies'' that AS might adopt in response to the large numbers of
actors, goals, activities, understandings and potential conflicts that
exist within large scale collaborations and the capabilities that
those AS need in order to implement those strategies. It also
considers two factors in producing an initial requirement
specification:– {\bf the complexity of the tasks and the complexity of
  the collaborations.} The former includes the nature of the goal
  relationships and the latter the nature of the collaborative
  relationships. Both factors are relevant to understanding and
  managing collaboration in virtual organisation. In this paper we
  report the capabilities and understandings that AS will require if
  they are to implement these ΄coping strategies‘. These strategies
  include mechanisms that address: 

  \begin{enumerate}
  \item {\bf Task Structure issues:} The large number of and variability in the goals, activities and state-descriptions inherent in large group collaboration must be identified, negotiated, monitored and judged. One possible solution is to have more abstract representations of each one that provide less variability (though loosing all aspects of that variability may not be desirable).   
  \item {\bf Group Structure issues:} The large and possible varied
    structure of the group(s) themselves. The issues of group
    structure and how that affects group awareness, group
    communication and group coordination. For example, one possible
    solution is appropriate division of the collaborative activity
    into subsections that can then be managed by addressing each
    subdivision only via a specified middle-manager or gate-keeper (of
    course, the consequence of this can be lack of awareness in the
    group and the gate keeper becomes the potential bottleneck).
  \end{enumerate}

\section{Large Collaboration Capability Requirements}
Previous research identified the capabilities needed to take part in
small-group collaborations \cite{4,5,10,11} as well as the characteristics
of large-groups \cite{3,6,15} that mean those capabilities are insufficient
for the participation in and management of large
collaborations. Previous work does not identify the collaborative
structures and processes needed to work as one or achieve a smooth
effortless ΄flow‘ when working in large groups. To address this, we
have developed a conceptual framework to allow us to develop the
detailed structures and processes needed for large collaborations in
highly dynamic situations of high emergence. We present an overview of
this in this paper for a more detailed expose see \cite{3}. 

\subsection{A Framework of Large-Group Collaboration}
In considering the collaboration requirements, we focused upon the
need for a collaborative group to be able to manage emergent
properties and dynamic changes to: 1. the organization (wider group)
within which a collaboration takes place, 2. the internal (sub-)
groups that undertake tasks within that collaboration, 3. the tasks
themselves, and 4. the resources required to undertake those tasks. As
we have already briefly mentioned collaboration and conflict are
inseparable, in that potential and actual conflicting situations arise
within collaborations. Hence designing collaborations without conflict
is impossible, instead we recognise that the collaborative structure
has to have conflict mechanisms.  We have identified three aspects of
conflict: a) the avoidance of conflict before it occurs, b) the
identification of conflict that cannot be avoided, c) the resolution
of the conflict identified in b, and, in the execution of those three
components the need for communication and coordination of the factors
identified above. The framework from which particular collaborative
capability requirements were identified is summarized in Figure 1,
below:  

\begin{figure}[ht]
\begin{center}
\includegraphics[width=\textwidth]{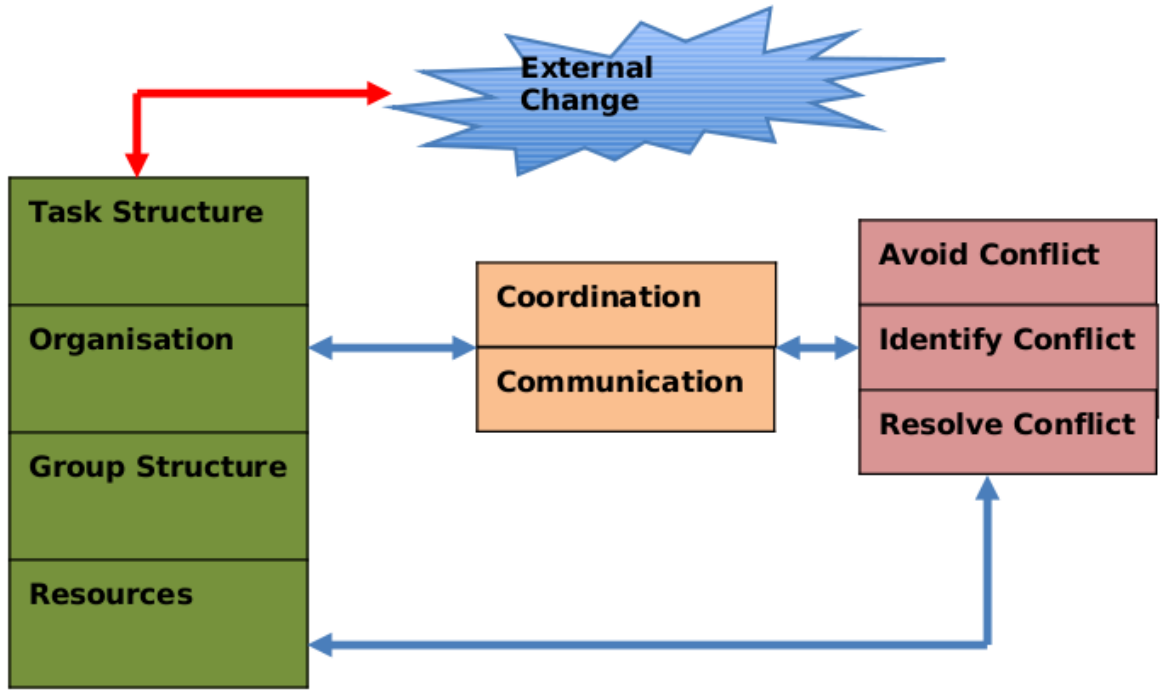}
\caption{Framework for large group collaborations.}
\label{fig:framework}
\end{center}
\end{figure}

\subsubsection{The Need for Small Group Collaboration Mechanisms}

The starting point for the large group collaboration framework is the
vast amount of previous research carried out on small group
collaborations (such as \cite{4,5,10,11,13}). Rather than offer this as an
alternative to those it should be seen as building upon them. Thus the
mechanisms required for large group collaboration are additional to
the requirements for small group collaboration moreover, small groups
can exist within large groups as well independently from
them. Consequently, both sets of mechanisms are needed and need to be
satisfied to engage in teamwork that has ``collaborative flow'' and
the ability to ``work as one''. The capabilities we have identified
that are required of small group collaboration are reported elsewhere
\cite{10,11,13} and will not be repeated here. 

\subsubsection{The Need for Large Group Collaboration Mechanisms}

As stated earlier, the large numbers of actors, goals, actions and resources that make up a large group collaboration place additional requirements for collaboration mechanisms to enable the achievement of collaborative flow in the context of the highly dynamic and emergent aspects of the required teamwork. Figure 2. Below situates the problem space for large groups collaboration (adapted from \cite{17}). Our focus is in P4, where there are problems of broad extent coupling diverse complex subsystems.  In P1, where we have problems of limited scope and limited complexity, simple coordination mechanisms will suffice. In P2, where we have problems of a limited extent but with high complexity, mechanisms for self-coordinating groups will suffice.  In P3, where we find problems of low complexity but high extent and diversity, structured collaboration mechanism are needed and will suffice.  

\begin{figure}[ht]
\begin{center}
\includegraphics*{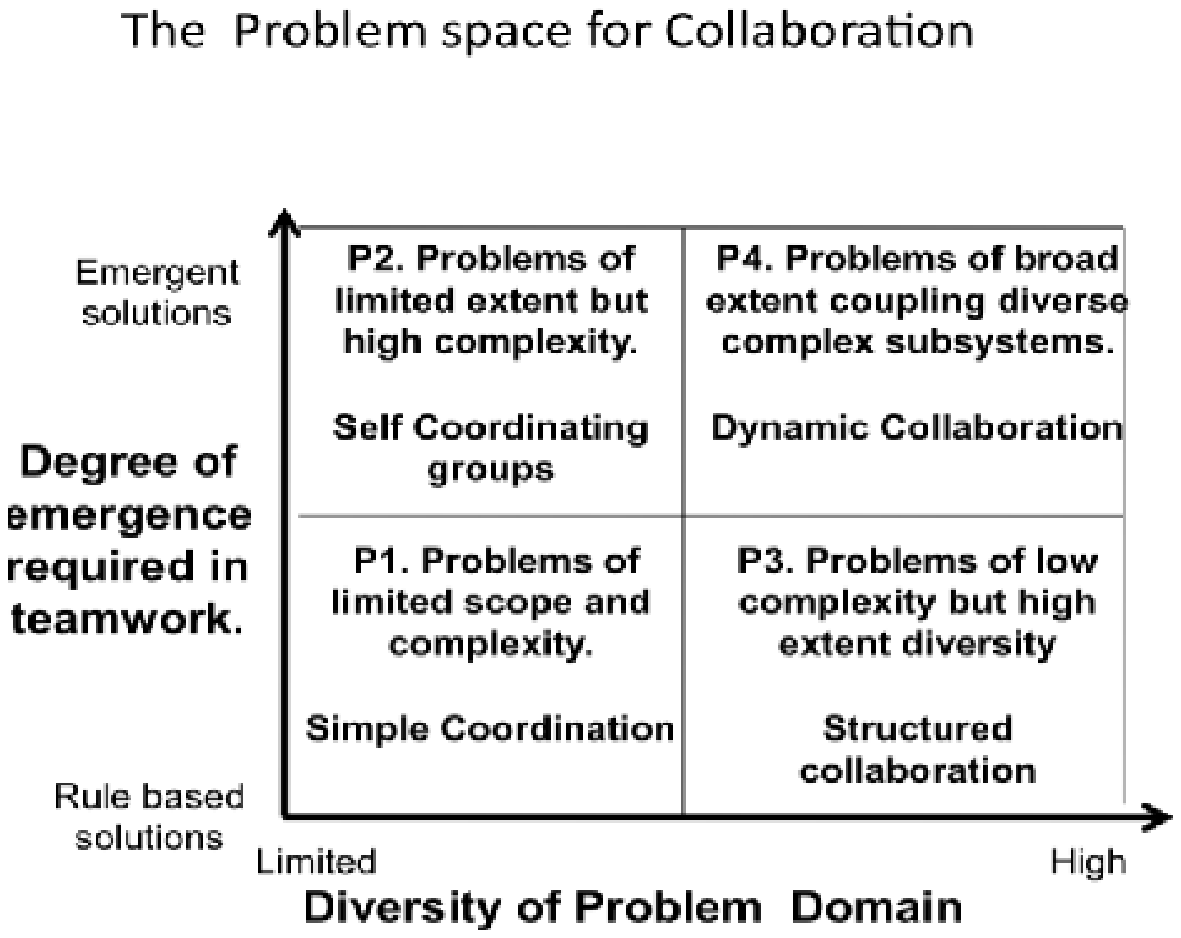}
\caption{The problem space for large group collaborations.}
\label{fig:problem}
\end{center}
\end{figure}

Hence, those involved in large collaborations will need further
capabilities and mechanisms in addition to those identified for small
group collaboration. More specifically for example, they will require
capabilities that will allow them to dynamically re-distribute
resources, dynamically share goals within groups of actors etc where
the extent and diversity of these are too large to allow 1:1
monitoring of each other’s action, sub-goal and outcome. Hence in
large group collaborations the importance of further mechanisms and
capability to address P4 in figure 2. 

\subsection{The Framework}

The Framework presented in Figure 1. describes a number of structures
and process relationships that come into play and influence
large-group collaborations.  There are a number of structures (on the
left-hand side of the framework and coloured green) covering task,
organisation, group and resources, where within each there are many
alternative structural relationships. For example, there are many
different types of task structures and, and within a given task
structure there will be different types of relationships between the
elements of the task. For example, within a task structure there may
be hierarchical goal structuring and a network structure for the
various procedures used. Similarly, there are many different types of
organisation structures and within a given organisational structure
there are different types of relations. Hence each of the task,
organisation, group and resource, structure cells of the framework
represent components that are themselves complex. Moreover, they are
dynamically changed by both internal and external factors and interact
with each other. Hence the different task group, organisation and
resource structures interact with each other to deliver a
collaborative mechanism. For further details of the various task,
group, organisation and resource structures see \cite{3}. The central cells
of the framework  (coloured orange in figure 1) capture the different
coordination and communication structures and processes that may exist
within a large-group collaboration.  The coordination structures and
processes operate across and within each task, group organisation and
resource structures. For example processes and structures for
coordinating resources relative to their use in tasks by groups
between organisations are detailed here. Similarly, communication
structures and processes operate to ensure that appropriate
information, understanding and awareness is achieved both about and
across the tasks, organisations groups and resources. Moreover,
particular communication structures and processes may take different
forms across the collaboration. For example the communication
processes within one sub-group may be strictly hierarchical, while in
another it may be possible for anyone to communicate with anyone and
everyone directly.  (As above, for further detail see \cite{3}). The right
hand-side of the framework (coloured pink in figure 1.) addresses the
processes and structures that are required in a collaboration to
avoid, identify and resolve conflict. This is explicit in the
framework because of the importance of conflict to collaboration.  The
three cells collectively capture the processes and structures managing
and offsetting conflict as it arises and before it arises that enable
resource conflicts, task conflicts, group conflicts, and
organisational conflicts to be overcome. Moreover, the relationships
between the cells on the green (left-hand side) and the pink cells
(right-hand side) of figure 1. are "piped‘ through the coordination
and communication processes and structures. Hence, the subject matter
of the coordination and communication processes relate to the
resolution, identification and avoidance of conflict relating to task,
group, organisational and resource properties. The need for these may
arise as a result of external factors forcing change or from internal
factors requiring adaption and change in one part resulting from
change in another part of the collaborative system. For the purposes
of this paper we describe the whole system driven from the perspective
of the right-hand (Conflict) side of figure 1, however we could
equally well describe the system starting from the left-hand (task,
group, organisation, resource) side.     

\subsection{Avoid Conflict}

Conflict is an integral part of collaboration \cite{14}; it is the management and reduction of conflict that leads to successful collaborations. Hence to understand collaboration one must consider conflict avoidance, identification and resolution. The first area described is the avoidance of conflict between the members of a large group and their understandings of the ways in which the task at hand would be achieved. Furthermore we also consider the roles, goals and actions to be adopted during the achievement of the task and the allocation of available resources in the course of that adoption. This section describes that first part of the framework.

\subsubsection{Task Structures -– Goals}

There are many different models of task structuring (indeed we have ourselves contributed to these, see for example, \cite{16}) however, while they may use different terms and have different intended uses they share a number of properties in common. Of concern here is the properties of collaboration structures and process required to achieve tasks, rather than the modelling or analysis of tasks themselves. 
The collaborating group must distribute the group’s work amongst its various actors and sub-groups. This distribution requires a set of capabilities of group members if it is to be achieved without external intervention. Those capabilities include the identification of local goals that will, when correctly scheduled and completed, lead to the achievement of the group’s, high-level goals, the mapping of those sub-goals to the achievement of high-level goals (i.e. an understanding of the relationship between the achievement of local goals within smaller sub-groups and the progress of the wider group towards its shared, higher level goals), the allocation of those sub-goals to appropriate actors and/or sub-groups and, where appropriate, the re-negotiation of that allocation with those actors and sub-groups.

\subsubsection{Task Structure -- Actions}

The actors and sub-groups must have and/or negotiate an understanding of the actions needed to achieve their local sub-goals. In some situations, this negotiation will require only an agreement that sub-goals will be achieved, in others that they will be achieved to a specified schedule (in line with the dependencies that exist between sub-groups and their local objectives). In other situations, the trust held by managers and leaders of the group will be so low that individual sub-groups maybe required to provide a detailed description of the way that sub-goals will be achieved, rather than being left to make their own way towards their own objectives, and hence gives rise to the need for further communication and coordination in the collaboration. 

\subsubsection{Organisational Structure}

The organisational structure refers to the pattern of relationships that exist both within a group(s) and in the organisation(s) in which the group(s) exists. For example a hierarchical group may exist within a coalition of services (as we find when military allies are created or when different care services work together). In homogeneous groups (i.e. groups with a single structure and in which each participant has an identical understanding of the group structure), certain conflicts would not arise i.e. the structure of the group would, by definition, be known by all group members. With heterogeneous groups participants would need to avoid conflict between competing understandings of the group structure to be adopted (e.g. what roles were needed and their definition, the process of allocating roles, the responsibilities of those roles etc). Participation in either ''homogenous'' and ``heterogeneous'' groups would require a number of key properties (i.e. collaboration requirements) to address external and internally driven dynamic changes: Structuring (and restructuring) of the group for appropriate resources (and resource changes), structuring (and restructuring) of the group to enable effective co-ordination and communications, structuring (and restructuring) to enable task (and changes to task) goals, structuring (and restructuring) of group  to meet external organisational needs (and organisation changes), - in all cases to avoid/mitigate conflicts arising in the collaboration.

\subsubsection{Group Structure - Roles}

Both hetero- and homogenous groups must agree the specific roles to be
played by individual actors in the course of a collaborative
activity. In a flat structure (i.e. one, in which each actor is able
to interact with and influence the work of each other actor), these
collaborative roles will differ very little from each other (i.e. the
information, requests and responses passed from one actor to another
will follow similar patterns, though the specific part of the
collaborative task undertaken may vary from actor to actor). 

In more structured groups, some actors will be asked to perform management roles (i.e. to direct the activity of other actors and / or to channel the information between the wider group and their ΄subordinates‘), some will be asked to subordinate themselves to such managers and some to adopt both management and subordinate roles (i.e. to become ΄middle-managers‘). The specific roles adopted within a particular collaboration will influence and be influenced by both the capabilities of the actors participating in a particular collaborative group and on the organizational structure of that group. Some groups of actors may, for example collaborate without the need for a central manager, some groups may have one or more managers imposed upon them by a higher authority and some may select managers by following a pre-determined algorithm such as voting amongst themselves. 
If correctly managed, the structuring of a large group through such an
allocation of roles will allow large groups to cope with the
impossibility of monitoring every actor, action and objective, whilst
ensuring that the group goals are achieved, available resources are
utilized appropriately, collaborative flow is maintained and the group
can adapt to dynamic changes such as the loss of an individual actor
the loss of a resource or the alteration of a high level goal by an
external authority.  

\subsubsection{Group Structure - Actors}

In either context, (i.e., homogeneous or heterogeneous organisational structures), the actors responsible for assembling a large, collaborative group of their peers would need to be able to identify the resources and capabilities needed for the completion of the task at hand (e.g. if the task involves the construction of a wall the need for a certain number of bricks, a quantity of mortar and abilities to both lay bricks and mix mortar). In more sophisticated cases, this capability may extend to the identification of different combinations of resources and capabilities, any one of which could be used to satisfy the requirements of the task (e.g. in the case of the wall-building example, either the bricks, mortar and construction capabilities, or an ability to transport an pre-fabricated wall from a storage location to the construction site). In such cases, the capability requirements demanded of any actor involved in group-assembly would then include the identification of appropriate resource/capability combinations, the mapping of specific combinations into collaborative task completion outcomes plus the identification of resources and capabilities available by potential group members.

\subsubsection{Resources}

The resources needed for the completion of tasks may be drawn from either local or global sources. In some cases, specific actors or sub-groups will hold resources. For example, an information resource may be available only to the sub-group, organisation, or location in which it is held. In other cases central resource stores may be appropriate – information may be made available to all members of the collaborative group, regardless of their small (sub) group membership. Similarly, a single fuel dump may be managed by a single sub-group and allocated to other group members in case of need.

\subsection{Identify Conflict}

Despite the best attempts of the participants involved in a specific collaboration dynamic changes (e.g. in group composition, group structure, overall or local goals, prevailing activity and available resources) will inevitably present the opportunity for unanticipated conflicts to arise. In order to resolve these unanticipated conflicts, actors will need to both maintain awareness of those different factors (from group composition to resource availability) and, as the next section makes clear, manage or resolve conflicts arising within and between them. 

\subsubsection{Maintain Awareness.}

In collaborating groups members must maintain awareness of the complex components of the prevailing collaboration if they are to first identify and subsequently address the potential for conflict.

{\bf Organisational and Group Structure \& Processes} They must, for example, maintain awareness of the organisational structure or structures that govern the prevailing collaboration. For example, in homogenous groups of AS that awareness can be relatively easily maintained, since each of the AS involved will, by definition have an identical understanding of group structure (i.e. they will all have an identical understanding of the hierarchical, holarchical or other group structure under which the group is operating). It should be noted that such common understanding cannot be assumed in heterogeneous groups.
 Even within homogeneous groups actors involved in large group collaborations must maintain some awareness of the composition of the group within which they are working. This does not mean that they must maintain understanding of every individual actor and group within the wider collaboration in which they are involved. They must be able to identify local managers, subordinates and contacts, must understand how communication can be achieved with each one and must also be aware of changes in role allocation that causes a new actors to be their manager, subordinate or contact:

{\bf Task Structure \& Processes} In any dynamic environment, a truly collaborative actor must also maintain awareness of the group’s goals and sub-goals. Once again, no comprehensive awareness of all such goals either can or must be maintained by a single actor. This limitation notwithstanding each actor in a large-group collaboration must maintain awareness of their own objectives and, depending on their role, those of managers, subordinates and/or contacts in other sub-groups. This understanding is important if actors are to coordinate their activity with others and, ultimately ensure that this activity contributes to the groups shared goals.
Each actor must also maintain awareness of their own actions and their effects on the goals of the other group members with which they are in contact (the managers, subordinates and contacts identified above). In large group collaborations, the wider effects of each action may not be understood by every member of the wider group but an effective group structure will ensure that an appropriate understanding will be available to those in key roles and, as a consequence, that collaborative flow is maintained as the group progresses towards its shared goals.

{\bf Resources.} Finally, actors must maintain awareness of the resources that they need to complete their activities. They must follow the extent to which they hold sufficient reserves of each resource (e.g. knowledge, fuel or payload) themselves, the amount of each resource that can be obtained within the sub-group to which they belong and, in case of need, the nature and amount of the resources that can be obtained from a central store. 

\subsubsection{Manage Dynamic Changes}
To exemplify the interactions and interconnectedness of the different components of the framework we consider how when a change occurs to one part it influences and affects everything else in the framework.
Though maintaining awareness of a group’s initial configuration and of
the changes to that collaboration are central to an actors
participation in collaborative work, it should be noted that this
awareness is not, in itself, sufficient for collaborative flow to be
maintained in many collaborations. In addition to understanding a
developing large-group collaboration, actor(s) must also be able to
adapt to (manage) dynamic changes within that collaboration. They
must, for example, be able to manage changes to the {\bf composition of the group} i.e. to deal with situations in which loss of functionality, changing priorities or instruction from a higher authority cause actor to either join or leave a group involved in collaborative activity.
In the homogenous, structured, large groups those joiners and leavers may have no immediate impact upon a particular actor (because the joining or leaving actor affects only a remote subgroup), may change the composition and therefore activity of the local sub-group or may lead to the replacement of an actors immediate superior, subordinate or contact within other sub-groups. 
This in turn may lead to dynamic changes to the {\bf task structure and processes} in the light of a new role being adopted by the actor. Those changes to group composition (or indeed other changes e.g. the loss of an important resource, a change in the environment within which the group is acting or fresh instructions from a remote authority) may in turn cause dynamic changes to local goals, high-level goals or both. 
In response to changes in group goals or composition, the actions of an individual actor may also need to be dynamically altered. The adoption of a new group structure, role or local (sub) goal will each cause a collaborative actor to reconsider their activity, the schedule to which that activity must adhere or both together.
Finally, the {\bf resources} needed by and available to an actor working within a large group are likely to change in the course of collaborative activity of any complexity. A lost communication channel can lead to the loss of information resources (i.e. those resources which were supplied by other actors), a blocked physical pathway may lead to lost fuel supplies and the loss of a superior or contact actors will prevent communication beyond the immediate sub-group within which an actor is operating. Adaptation to such losses may only be possible if actors posses the capabilities to: a) dynamically change goals, group structure and activity in light of changes to resource needs and b) dynamically change goals, group structure and activity in light of changes to resource availability.

\subsection{Address \& Resolve Conflict}

When conflicts are identified in the course of a large-scale collaboration actors will also need the capability to address and resolve them. In perhaps the least disruptive case, the actor(s) identifying a conflict may also be able to resolve it. This may require the revision or change of individual tasks (which in-turn may impact upon the wider task of the group). It may also require an alternative or additional resource usage (e.g. the allocation of more time to a transport task in exchange for a lower fuel usage). To the extent that these changes are made they must be done so with communication and coordination to the appropriate other members of the collaboration to ensure that awareness and the potential for further conflict is minimised. Hence we see a further instance of the complex interactions that occur with the collaboration framework

\subsection{Communication and Coordination Structures and Processes}

In describing the framework here we have left largely implicit the
detail of the communication and coordination structures and processes
needed. Briefly, the coordination structures and processes
characterise the dependencies and the means for ensuring the required
states between those dependencies are maintained. These exist both in
the individual cells of the framework and between the cells of the
framework. For example within the task structures there will be
coordination processes needed to ensure that the completion of tasks
are coordinated within the group. Furthermore, between the resource
and task structures there will be coordination process to ensure the
availability of resources in timely manner.  
Similarly, the communication processes and structures are the rules
governing the routes and the form of communication within the
collaboration. These may be universal (i.e. one set of rules applies
to all) or may be diverse (i.e. different rules apply to different
parts) and static or dynamic (i.e. they may be allowed to change
over-time and/or events or not). These communication rules will again
have implications and influences upon the group, task, organisation
and resource structures and upon how conflict is avoided, identified
and resolved in collaboration. (For a fuller description see \cite{3}). 

\section{APPLICATION TO AUTONOMOUS SYSTEMS and Beyond}

We have briefly described the collaboration framework from the
perspective of the ``Identify conflict'', ``Avoid conflict'' and
΄Resolve Conflict‘ sections of Figure 1 and considered some of the
capabilities that collaborative groups of people and/or autonomous
systems will need. Future work will develop simulations of autonomous
systems that implement those capabilities and will then use those
simulations as the basis for testing the validity and practicality of
those candidate requirements. The research will also apply the
collaboration framework to heterogeneous groups of AS to consider: 
\begin{itemize}

\item different communication pathways between AS in different parts of the wider collaboration,
\item different understandings of the division of labor between different parts of the group,
\item different allocations of resource both to individual AS and the sub-groups to which they belong,
'item different coping strategies in the case that conflicts arise, such as those described above.
\end{itemize}

\subsection{Further Issues}

In applying the framework to AS we recognise there maybe collaboration problems arising that cannot be resolved by the AS encountering the problem. This may occur because AS have not been designed to play a full collaborative part in an activity. Alternatively, AS may incur damage, resources run out and group members lose contact with the wider group. In both cases this may affect individual and group ability to;
\begin{itemize}
\item avoid group structure, group composition, role, goal, activity and resource conflicts,
\item identify group structure, group composition, role, goal, activity and resource conflicts and
\item resolve group structure, group composition, role, goal, activity
  and resource conflicts.
\end{itemize}

Moreover, this may lead to a situation:
\begin{itemize}
\item where a partial or complete inability to {\it avoid} conflicts
  will result in an increased need to {\it identify} and {\it resolve} conflicts, and
\item a partial or complete inability to {\it identify} conflicts will
  obscure any remaining capability an AS holds to {\it resolve} those same
  conflicts.
\end{itemize}

In such situations, we may use the framework to identify a fallback
solutions or ``work-around'' that can be used to improve large group AS
functionality (where a full collaborative capability is not
available), and to identify the collaboration costs and limitations
associated with each work-around.

\subsection{Relevance to Virtual Organisations}

The framework we have presented here is of a conceptual nature, largely of relevance to those interested in understanding and/or developing collaborative organisations. It provides a basis for developing detailed models of the interactions that go on in collaboration and for modelling the structures and processes in a collaborative organisation. An important area is the development of mechanisms and technologies to support such collaborative organisations that can (and must) be developed following this framework. Research we are engaged in beyond our AS work includes work with health organisations and with local authorities where we are helping them to engage in collaborative decision-making, and collaborative service provision. In many cases these collaborations are creating new virtual organisations (VO) that come together to deliver and develop services, and which involve people from many different individual organisations and groups. To support these applications we need to develop mechanisms and technologies that address: 1. the development of a VO, 2. the functioning of a VO, and 3. the assessment of the collaboration in a VO. We envisage the development of collaborative VOs using the framework to also require tools and languages to allow the proposed structures to be expressed and reasoned about as an aid to design. To support the functioning of the VO we envisage that environments and tools that support easy and efficient sharing of information, formation of policy, decision-making and communication and coordination will be needed. While to support assessment of collaboration we will need to develop metrics of such aspects as the amount of communication or ease of communication, the amount of consensus, and sharing that exists within the organisation and the ability to avoid, identify and resolve conflict. All of these require language and software technologies that can carry with them a change of culture that allows organisations to work collaboratively to meet the demands of complex dynamic situations.  

\section{Conclusions}

In conclusion, therefore, this paper extends our understanding of
collaborative structures and processes. It situates large group
collaboration within the broader context of social emergence \cite{15}. It
has led to a set of requirements for future generations of autonomous
systems capable of participating in collaborative activity. We also
identify areas in which future research must extend the work presented
here. More specifically, those extensions must include:

\begin{itemize}
\item Investigation of AS reaction to ``individually-unsolvable'' role, goal, action and resource conflict,
\item Testing of both the validity and utility of the requirements identified here,
\item Application to human collaborations in service delivery.

\end{itemize}
Development of technologies to support dynamic collaboration in large-scale groups.
These extensions are the subject of our further research. 

\section*{Acknowledgements}

  Some of the work reported in this paper was funded by the Systems
  Engineering for Autonomous Systems (SEAS) Defence Technology Centre
  established by the UK Ministry of Defence.

\bibliographystyle{eptcs}
\bibliography{johnson}

\end{document}